\documentclass[11pt,a4paper]{article} 
\usepackage{jheppub}

\usepackage{graphicx}
\usepackage{amsmath}
\usepackage{amsfonts}
\newcommand{\beq}{\begin{equation}}
\newcommand{\eeq}{\end{equation}}
\newcommand{\beqa}{\begin{eqnarray}}
\newcommand{\eeqa}{\end{eqnarray}}
\def\lsim{\raise0.3ex\hbox{$<$\kern-0.75em\raise-1.1ex\hbox{$\sim$}}}
\def\gsim{\raise0.3ex\hbox{$>$\kern-0.75em\raise-1.1ex\hbox{$\sim$}}}

\def\x{{\bf x}}

\def\0{{\bf 0}}

\def\cal{\mathcal}

\title{Heavy-flavour production in high-energy d-Au and p-Pb collisions}

\author[a]{Andrea Beraudo}
\author[a]{Arturo De Pace}
\author[a]{Marco Monteno}
\author[a]{Marzia Nardi}
\author[a]{Francesco Prino}

\affiliation[a]{INFN, Sezione di Torino, via Pietro Giuria 1, I-10125 Torino}

\emailAdd{beraudo@to.infn.it}
\emailAdd{depace@to.infn.it}
\emailAdd{monteno@to.infn.it}
\emailAdd{nardi@to.infn.it}
\emailAdd{prino@to.infn.it}

\abstract{Soft-hadron measurements in high-energy collisions of small systems like p-Pb and d-Au show peculiar qualitative features (long-range rapidity correlations, flattening of the $p_T$-spectra with increasing hadron mass and centrality, non-vanishing Fourier harmonics in the azimuthal particle distributions) suggestive of the formation of a strongly-interacting medium displaying a collective behaviour, with a hydrodynamic flow as a response to the pressure gradients in the initial conditions. Hard observables (high-$p_T$ jet and hadron spectra) on the other hand, within the current large systematic uncertainties, appear only midly modified with the respect to the benchmark case of minimum-bias p-p collisions. What should one expect for heavy-flavour particles, initially produced in hard processes but tending, in the nucleus-nucleus case, to approach kinetic equilibrium with the rest of the medium? This is the issue we address in the present study, showing how the current experimental findings are compatible with a picture in which the formation of a hot medium even in proton-nucleus collisions modifies the propagation and hadronization of heavy-flavour particles.}

\begin{document}

\maketitle

\section{Introduction}
One of the most surprising findings in the experimental search for the Quark-Gluon Plasma is certainly the signature of possible collective effects, suggestive of the formation of a hot strongly-interacting medium, recently observed in the collisions of small systems like p-Pb at the LHC and d-Au (and now also $^3$He-Au) at RHIC, in particular when selecting events characterized by a high multiplicity of produced particles. Various observables support the above picture: the structure of two-particle correlations in the $\Delta\eta\!-\!\Delta\phi$ plane (double ridge), suggestive of a boost-invariant initial condition, with azimuthal spatial asymmetries mapped by the strong interactions into the final particle spectra~\cite{Abelev:2012ola,Aad:2012gla,CMS:2012qk,Adare:2013piz}; the non-vanishing values of the elliptic, triangular and higher flow-harmonics~\cite{Adare:2015ctn,Aad:2014lta}, obtained also through the study of higher-order cumulants~\cite{Chatrchyan:2013nka}, which seem to indicate a common correlation of all the particles with the same symmetry-plane; the hardening of the $p_T$-spectra moving towards more central events~\cite{Abelev:2013haa,Chatrchyan:2013eya}, which can be described as the effect of the collective radial flow of an expanding medium. The above effects display also a characteristic dependence on the particle species (mass ordering), still in agreement with the expectations of a hydrodynamic description~\cite{ABELEV:2013wsa,Khachatryan:2014jra,Adare:2014keg}. Theory predictions based on such a picture can be found e.g. in Refs.~\cite{Bozek:2011if,Bozek:2013uha,Bozek:2013ska,Werner:2013tya,Romatschke:2015gxa}.

The possible formation of a medium featuring a collective behaviour in proton-nucleus collisions, on the other hand, does not seem to significantly affect the yield and momentum distribution of hard observables like jets and high-$p_T$ hadrons: minor changes with respect to the p-p case (after accounting for the proper scaling with the number of binary nucleon-nucleon collisions) can be attributed to initial-state effects, like the nuclear modification of the Parton Distribution Functions (PDF's). The nuclear modifications factor $R_{\rm pPb}$ of jets~\cite{ATLAS:2014cpa,Adam:2015hoa} and charged hadrons~\cite{Abelev:2014dsa} at the LHC was found to be compatible with unity within the experimental error bars, although these findings have to be taken with a grain of salt, due to the absence of a p-p benchmark at the same center-of-mass energy which introduces large systematic uncertainties: alternative interpolations of p-p data taken at lower and higher energies can lead to different results~\cite{Khachatryan:2015xaa}. These findings are not necessarily in contradiction with what observed in the soft sector, since in-medium parton energy-loss -- playing a major role in heavy-ion (A-A) collisions -- has a strong dependence on the the temperature and the size of the medium (for the average energy-loss due to coherent gluon radiation one has $\langle\Delta E\rangle\!\sim\!\hat{q}L^2\!\sim\!T^3L^2$~\cite{Baier:1996kr}), while on the contrary for the radial velocity of the medium at time $t$ one finds the approximate behaviour $v^{x/y}\!\sim\!c_s^2t/\sigma_{x/y}$~\cite{Ollitrault:2008zz}: in this case the speed of sound $c_s$ has only a mild temperature dependence and the smaller transverse size $\sigma_{x/y}$ of the system in p-A with respect to A-A leads to larger pressure gradients and hence to a larger radial flow.

In light of the above findings it is clearly of interest what happens to heavy flavour (HF) particles in such small systems: in fact, due to their large mass, the initial charm and beauty production occurs in hard pQCD processes on a very short time-scale, like all other hard particles; at the same time, however, experimental data show that in A-A collisions they tend to acquire at least part of the elliptic and radial flow of the medium. Is the hot medium possibly formed in p-A collisions, although of  small size and short lifetime, able to leave its signatures in the final hadronic observables also in the heavy-flavour sector? For several years the paradigm was that such collisions were only useful to point out initial-state effects, like nuclear modifications of the PDF's. First experimental data, due to their large systematic uncertainties, couldn't rule out such an interpretation arising from the above theoretical prejudice. However, as more experimental data are getting accessible, it is clearly of interest to use them to discriminate among the different theoretical scenarios, including or not final-state effects.
The current experimental situations is the following. Results for the nuclear modification factor of the spectra of various heavy-flavour particles have been obtained for different colliding systems and center-of-mass energies: non-photonic electrons (NPE's) from charm and beauty-hadron decays in d-Au collisions at $\sqrt{s_{\rm NN}}\!=\!200$ GeV by PHENIX~\cite{Adare:2012yxa}, D mesons and HF electrons by ALICE~\cite{Abelev:2014hha,Adam:2015qda}, high-$p_T$ B mesons and b-jets by CMS~\cite{Khachatryan:2015uja,Khachatryan:2015sva} and $J/\psi$'s from B decays by LHCb~\cite{Aaij:2013zxa} in p-Pb collisions at $\sqrt{s_{\rm NN}}\!=\!2.76$ TeV. Overall, one can claim that -- within the large systematic uncertainties -- the measured $R_{\rm pPb}$/$R_{\rm dAu}$ is compatible with unity, i.e. no final-state effects, although PHENIX data lie systematically above unity.
Besides the $p_T$-spectra, it is clearly of interest to study how HF particles are distributed in the azimuthal plane and how they are correlated among themselves and with the other particles produced in the collision, with a double purpose: both for looking for possible medium-modifications of the original $Q\!-\!\overline{Q}$ correlations from the initial hard production and for checking whether correlations with the other hadrons reflect a common correlation with the same symmetry plane characterizing the initial condition, i.e. whether also HF particles tend to follow the flow of the medium (in case the latter is created). 
Due to the small branching ratio ($\sim$4\% for $D^0\!\to\!K\pi$) a direct measurement of $D\!-\!\overline{D}$ correlations is currently out of reach. Preliminary results from ALICE for D-h and e-h correlations in p-Pb collisions are available~\cite{Bjelogrlic:2014kia,Filho:2014vba}. However, if the purpose is to display the possible angular decorrelation of the original $Q\overline{Q}$ pairs from the hard event, the comparison with theoretical calculations may result difficult, since part of the away-side hadrons can come from the fragmentation of light jets from NLO contributions to heavy-quark production, like flavour-excitation or gluon-splitting processes. A more direct link with the parent heavy-quarks is provided by PHENIX measurements of e-$\mu$ correlations~\cite{Adare:2013xlp}, both leptons (after background subtraction) having a heavy quark as an ancestor; comparing p-p and d-Au measurements one observes a suppression of the away-side yields, suggesting a decorrelation due to the interaction with a medium.
On the other hand, as above mentioned, correlations of heavy-flavour particles with the other charged hadrons from a given event may be studied in order to check whether they display the same long-range structure and azimuthal modulation observed in the soft sector and interpreted as arising from the elliptic (and possible higher harmonics) flow of a strongly-interacting medium. Besides the above mentioned preliminary study of e-h correlations~\cite{Filho:2014vba}, the ALICE collaboration has recently presented results for forward-central $\mu$-h correlations in p-Pb collisions~\cite{Adam:2015bka}. In central events, if the jet-like contribution estimated from peripheral collisions is subtracted, a double ridge structure reminiscent of what found for light hadrons appears: this can be interpreted as a signal of elliptic flow of the muons, part of which coming from charm and beauty decays.

To summarize: even if strong conclusions cannot be drawn, there are hints (although not an evidence) from recent experimental data that heavy-flavour particles produced in the collisions of small systems may be characterized by a finite elliptic flow, signature of the rescattering with a strongly-interacting medium. Can this lead one to reconsider the interpretation of the results for the nuclear modification factor? Can the statement that the latter looks compatible with unity arise from the theoretical prejudice that no medium can be produced in p-A or d-A events and, in case it were produced, it would lead to a quenching of the spectra?
An interesting analysis was carried out in Ref.~\cite{Sickles:2013yna}, where blast-wave spectra for D and B mesons -- with parameters fixed using light hadron spectra -- turned out to be able to explain the $R_{\rm dAu}\lsim 1.5$ of the HF decay electrons in d-Au collisions at $\sqrt{s_{\rm NN}}\!=\!200$ GeV, the value larger than unity at moderate $p_T$ being interpreted as due to the radial flow acquired in the medium. Although the above interpretation looks suggestive, it it nevertheless based on a quite simplified picture, assuming that heavy mesons follow the flow of the other hadrons, without wondering whether the latter is a realistic assumption.
Here we wish to provide a more solid theoretical framework to study HF observables in the collisions of small systems, focusing on the d-Au and p-Pb cases at RHIC and LHC.
Here, a proper hydrodynamic background for the heavy-quark propagation is developed (some preliminary results were shown in~\cite{Nardi:2015pca,Beraudo:2015vja}). Initial-state nuclear effects (nPDF's and transverse-momentum broadening) are included in the initial hard production of the $Q\overline{Q}$ pairs, which are then distributed in the transverse plane according to the local density of binary collisions. Assuming that, within a quite short time interval $\tau_0$, a thermalized medium is formed, which will live for about 2-3 fm/c in the deconfined phase, one can follow the evolution of charm and beauty quarks in such a hydrodynamic background by solving the relativistic Langevin equation developed in Refs.~\cite{Beraudo:2009pe,Alberico:2011zy,Alberico:2013bza}. Finally, heavy quarks are hadronized according to a mechanism~\cite{Beraudo:2014boa} which involves their recombination with light partons from the medium to form colour-singlet objects (strings), eventually fragmented to produce the final hadrons. The results of the above setup, accounting for the heavy-quark propagation and hadronization in the presence of a hot deconfined medium, is that final HF particles (D and B mesons and their decay electrons) are characterized by a non-vanishing radial and elliptic flow, providing a new paradigm to interpret current experimental data.

\section{The setup}
In A-A collisions the main source of initial eccentricity (at least for what concerns elliptical deformations) is represented by the finite impact parameter of the two nuclei. In this case, for the initialization of the hydrodynamic evolution of the medium, smooth average initial conditions based on the optical Glauber model are sufficient to describe the gross features of the system, which tend to develop an elliptic-flow as a response to the initial spatial anisotropy. Hence, in our previous studies focused on Au-Au and Pb-Pb collisions~\cite{Alberico:2011zy,Alberico:2013bza,Beraudo:2014boa}, we could rely on such a picture. The situation gets different already when one starts looking at different observables, like elliptic flow in ultra-central events or triangular flow, which can only arise from event-by-event fluctuations, not captured by the optical Glauber model. It is reasonable to assume that the major source (although not the only one) of fluctuations in the A-A case is represented by the random positions of the nucleons inside the colliding nuclei in each event. Such fluctuations can be easily simulated by standard Monte-Carlo (MC) implementations of the Glauber model~\cite{Miller:2007ri}, in which randomly distributed nucleons of the two different nuclei collide if their transverse distance $d$ is such that $d\!<\!\sqrt{\sigma_{\rm NN}^{\rm in}/\pi}$ ($\sigma_{\rm NN}^{\rm in}$ being the inelastic nucleon-nucleon cross-section).
If the necessity of a Glauber-MC approach in the A-A case was realized only when addressing observables ignored in the first experimental studies, in the collisions of small systems (d-Au, p-Pb and also, recently, $^3$He-Au) it is absolutely mandatory: initial-state anisotropies have little to do with the value of the impact parameter and are instead dominated by event-by-event fluctuations in the nucleon positions and possibly, at a more microscopic scale, in the colour-fields generated by the valence partons of the colliding hadrons. {Although there are attempts in the literature to account for these sub-nucleonic fluctuations in the initialization of hydrodynamic equations~\cite{Bzdak:2013zma}, we neglected them, adopting a much simpler approach}.

\begin{table}[h]
\begin{center}
\begin{tabular}{|c|c|c|c|c|}
\hline
System & $\sqrt{s_{\rm NN}}$ & $K\tau_0$ & $\tau_0$ (fm/c) & $\sigma_{\rm smear}$ (fm)\\
\hline
d-Au & 200 GeV & 6.37 & 0.25 & 0.2-0.4\\
\hline
p-Pb & 5.02 TeV & 3.99 & 0.25 & 0.2-0.4\\
\hline
\end{tabular}
\end{center}
\caption{The parameter set used to initialize the hydrodynamic equations.}
\label{tab:init}
\end{table}
As above mentioned, for the simulation of the initial conditions of the hydrodynamic evolution of the medium -- which will represent the background in which the heavy quarks will be eventually distributed and made propagate -- we relied on a Glauber-MC model. Each binary nucleon-nucleon collision was assumed to deposit some entropy in the transverse plane, described by a Gaussian distribution centered around the scattering position and depending on the smearing parameter $\sigma_{\rm smear}$. This leads to the initial entropy-density profile
\beq
s(\x)=\frac{K}{2\pi\sigma_{\rm smear}^2}\sum_{i=1}^{N_{\rm coll}}\exp\left[-\frac{(\x-\x_i)^2}{2\sigma_{\rm smear}^2}\right].\label{eq:sdens}
\eeq 
In the above $K$ is a constant, fixed linking $K\tau_0$ ($\tau_0$ being the thermalization time at which the hydrodynamic evolution starts) to the final rapidity density $dN_{\rm ch}/d\eta$ of charged hadrons (taken as a proxy of the initial entropy), according to the procedure described in Appendix~\ref{app:init}. The full set of parameters employed in the estimate of the initial entropy distribution is given in Table~\ref{tab:init}.
Due to the random location of the participant nucleons and hence of the binary collisions, the above initial condition is characterized by an elliptic deformation quantified by the eccentricity $\epsilon_2$ (a positive real number) and by its azimuthal orientation $\Psi_2$~\cite{Alver:2008zza}:
\beq
\epsilon_2\,e^{i2\Psi_2}\equiv-\frac{\left\{r^2\,e^{i2\phi}\right\}}{\left\{r^2\right\}},
\eeq
where the curly brackets denote an average over the transverse plane weighted by the entropy density in Eq.~(\ref{eq:sdens}).
The eccentricity $\epsilon_2$ is then given by
\beq
\epsilon_2=\frac{\sqrt{\{y^2-x^2\}^2+4\{xy\}^2}}{\{x^2+y^2\}}
\eeq
and the event-plane angle\footnote{Usually the term \emph{event-plane} refers to the estimate of $\Psi_2$ from the final hadron distribution.} $\Psi_2\in[-\pi/2,\pi/2]$ by
\beq
\Psi_2=\frac{1}{2}\,{\rm atan2}\left(-2\{xy\},\{y^2-x^2\}\right).
\eeq
The eccentricity $\epsilon_2$ coincides with the usual \emph{reaction-plane} eccentricity if one rotates the nucleon positions by the angle $\Psi_2$, so that the minor axis of the ellipse is aligned along the $x$-axis: 
\beq
\epsilon_2=\epsilon_{\rm 2,rot}^{\rm RP}=\frac{\{y^2-x^2\}_{\rm rot}}{\{y^2+x^2\}_{\rm rot}}.
\eeq

\begin{table}[h]
\begin{center}
\begin{tabular}{|c|c|c|c|c|c|}
\hline
Nucleus & $A$ & $R$ (fm) & $\delta$ (fm) & $\rho_0$ (fm$^{-3}$) & $\sigma_{\rm NN}^{\rm in}$ (mb)\\
\hline
Au & 197 & 6.38  & 0.535 & 0.1693 & 42\\
\hline
Pb & 208 & 6.62 & 0.546 & 0.1604 & 70\\
\hline
\end{tabular}
\end{center}
\caption{The parameters used in the Glauber-MC modeling of the collision.}
\label{tab:woods}
\end{table}
To initialize the hydrodynamic evolution we proceeded then as follows. A few thousands nuclear configurations were generated, distributing the nucleons randomly according to the Fermi distribution
\beq
\rho(r)=\frac{\rho}{e^{(r-R)/\delta}+1},
\eeq
with parameters given in Table~\ref{tab:woods}. In the deuteron case, the relative position of the two nucleons was taken from a Hulten wave-function. A random impact parameter was then extracted from a distribution $dP(b)\!\sim\!b\,db$. In order to increase the statistics without storing too much information, a fixed number of trials was made for each configuration and the event was kept if at least one binary nucleon-nucleon collision occurred. At the end we generated 8520 and 8260 minimum bias collisions in the d-Au and p-Pb cases, respectively, each one characterized by the entropy distribution in the transverse plane given by Eq.~(\ref{eq:sdens}). According to the number of binary collisions events were divided in centrality classes.
Notice that in reality, depending on the their impact parameter and on the fluctuations of colour sources, each nucleon-nucleon collision can provide a different contribution (often modeled in the literature through a negative-binomial distribution) to the final particle multiplicity $N_{\rm ch}$: experimentally, events are classified according to $N_{\rm ch}$, rather than  $N_{\rm coll}$.   
For this work, we treated all nucleon-nucleon collisions on equal footing, neglecting such a further source of fluctuations.

\begin{figure}[!t]
\begin{center}
\includegraphics[clip,width=0.48\textwidth]{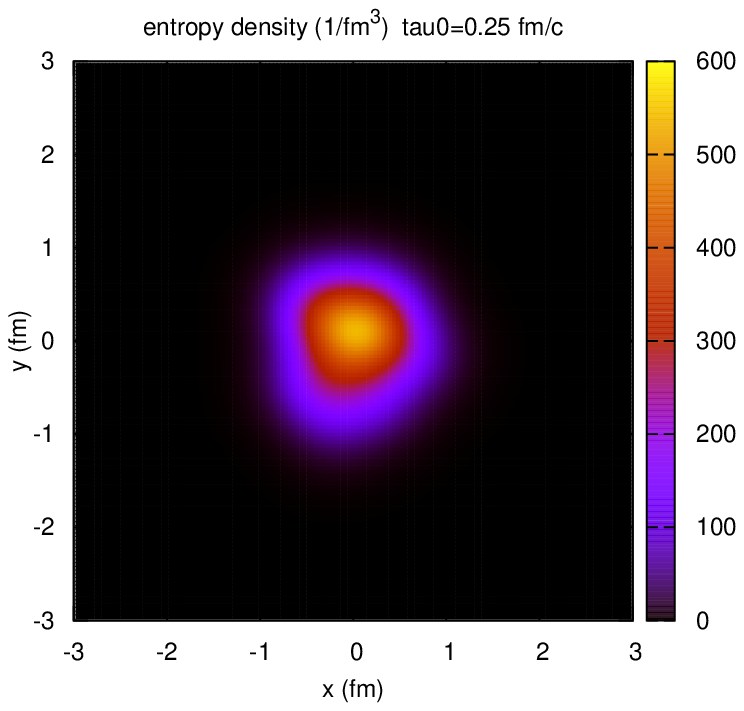}
\includegraphics[clip,width=0.48\textwidth]{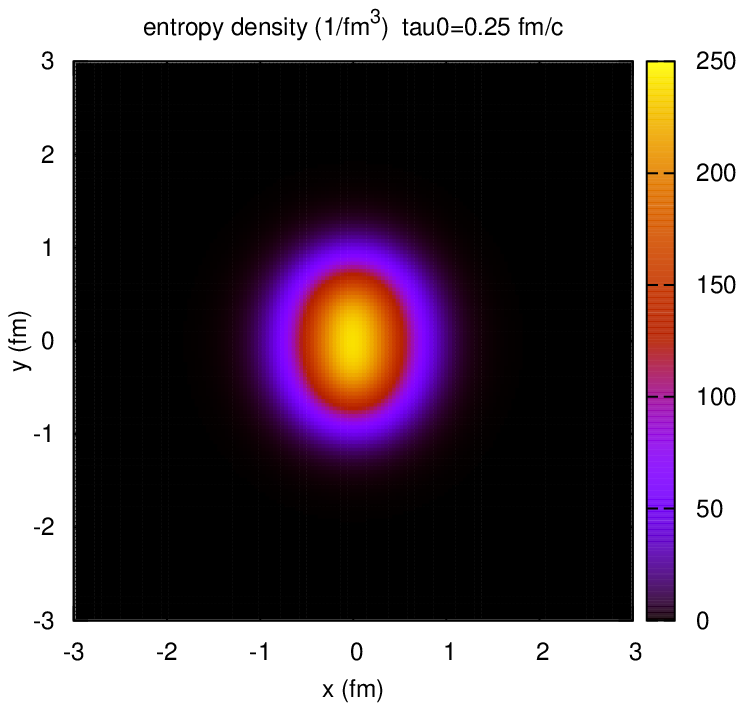}
\caption{The initial entropy-density profile for a single ultra-central p-Pb collision at $\sqrt{s_{\rm NN}}\!=\!5.02$ TeV provided by our Glauber-MC simulation (left panel) and the result of the weighted average of the 0-20\% most central events (right panel).}\label{fig:init} 
\end{center}
\end{figure}
\begin{figure}[!h]
\begin{center}
\includegraphics[clip,width=0.48\textwidth]{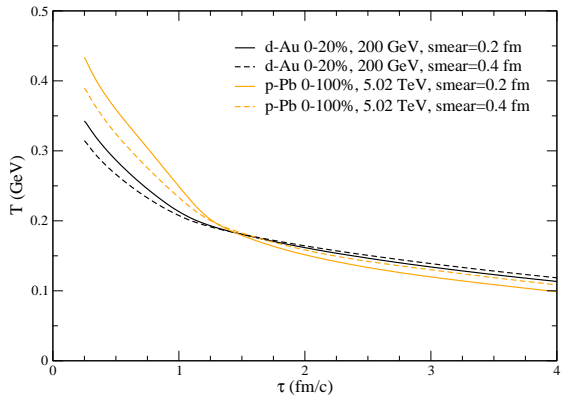}
\caption{The temperature evolution $T(\tau,\x\!=\!\0)$ of the medium formed in d-Au and p-Pb collisions, for various values of the gaussian smearing. The curves refer to the centrality classes so far employed for heavy-flavour experimental analysis in small systems.}\label{fig:temperature} 
\end{center}
\end{figure}

Full event-by-event hydro+transport simulations of HF production would require huge storage and computing resources. For this first study we decided to adopted a simplified approach. For each centrality class considered in our analysis defined as a fraction (percentile) of the total cross section (e.g. 0-20\%, 0-100\%) we took the average of all the events belonging that class, each one weighted by its value of $N_{\rm coll}$ (since we wish to have an average backround for HF propagation, whose production is a hard process scaling with the number of binary collisions) and rotated by $\Psi_2$.
Fig.~\ref{fig:init} shows the result of such a procedure, comparing the initial profile of a single central p-Pb collision -- with a quite irregular shape -- to the average one for the 0-20\% centrality class, the latter being characterized by a clear elliptic eccentricity.
The smooth entropy-density profile thus obtained was used as the initial condition of the hydrodynamic evolution, calculated with the ECHO-QGP code~\cite{DelZanna:2013eua}. We performed viscous runs with $\eta/s\!=\!0.08$, corresponding to the universal lower bound predicted by the gauge-gravity duality.  Concerning the longitudinal direction, in order to employ a reasonable amount of computing resources for this first analysis, we took a rapidity-flat profile with a Bjorken-like flow $v^z=z/t$, reducing the calculation of the hydrodynamic background to a (2+1)D problem. The resulting temperature evolution of the medium is displayed in Fig.~\ref{fig:temperature}.

\begin{figure}[!h]
\begin{center}
\includegraphics[clip,width=0.48\textwidth]{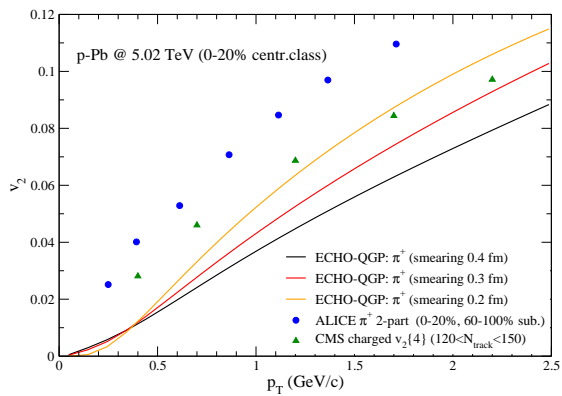}
\caption{The elliptic-flow of charged pions in the 20\% most central (highest $N_{\rm coll}$ in our setup) p-Pb events at $\sqrt{s_{\rm NN}}\!=\!5.02$ TeV from the hydrodynamic evolution of our average initial condition, with different smearing parameters. Also shown, for comparison, are ALICE~\cite{ABELEV:2013wsa} and CMS~\cite{Chatrchyan:2013nka} data obtained with 2 and 4-particle correlations.}\label{fig:pionv2} 
\end{center}
\end{figure}
In spite of the above simplifications, we believe to have a sufficiently realistic background for our purposes and, in particular, {to be} able to provide predictions for the possible elliptic flow acquired by heavy-flavour particles.
{This might be of relevance in order to attempt an interpretation of the recent electron-hadron and muon-hadron correlations measured in p-Pb collisions}.
As a validation of our hydrodynamic background, in Fig.~\ref{fig:pionv2} we compare the outcomes for the pion elliptic-flow in central p-Pb collisions arising from our calculations (with different smearing parameters) with the data obtained by the ALICE~\cite{ABELEV:2013wsa} and CMS~\cite{Chatrchyan:2013nka} collaborations (the 4-particle cumulant analysis by CMS should remove non-flow effects). {Notice also that the two measurements, based on two and four-particle correlations, differ in the sensitivity to flow fluctuations, tending to overestimate/underestimate the actual magnitude of the average $v_2$, respectively.}
The size of the effect is {approximately} reproduced;
The decrease of $v_2$ with the increase of the smearing parameter can be understood as a consequence of the lower initial eccentricity obtained with larger values of $\sigma_{\rm smear}$.
We postpone a full event-by-event analysis ({clearly desirable, due to the importance of fluctuations in such small systems}), with a (3+1)D hydrodynamic evolution of the medium, to a future publication.

Having settled the background, both in the case of p-Pb and d-Au collisions, we can now address the initial $Q\overline{Q}$ production, which is simulated through the POWHEG-BOX package~\cite{Alioli:2010xd}. As in our previous studies, EPS09~\cite{Eskola:2009uj} nuclear modifications of the PDF's are adopted for the Au and Pb nuclei (only the central value is employed); on the contrary, no correction is used for the deuteron projectile. Within the POWHEG-BOX framework the calculation of the hard event is interfaced to PYTHIA, which takes care of the simulation of other processes such as initial and final-state radiation and intrinsic-$k_T$. In the p-Pb and d-Au cases heavy quarks are assigned a further transverse-momentum broadening proportional to the average number of binary nucleon-nucleon collisions $\langle N_{\rm coll}\rangle$ in the considered centrality class. 
For such an additional $k_T$-broadening in p-A events one has, $\Delta^2$ being the average squared-momentum acquired by the incoming partons of the proton in each individual nucleon-nucleon collision~\cite{Vogt:2001nh,Thews:2005vj},
\beq
\langle k_T^2\rangle_{\rm pA}=(1/4)\Delta^2\langle N_{\rm coll}\rangle,
\eeq
since, on average, one-half of the collisions will occur before the hard scattering and the transverse momentum broadening acquired by the initial-state parton will be shared by the quark and the antiquark of the heavy pair; in the d-A case the above estimate is further divided by a factor 2, since the collisions involve with equal probability both nucleons of the deuteron. Heavy quarks are then distributed in the transverse plane according to the local entropy density $s(\tau_0,\x)$.

Heavy-quark propagation in the QGP (assuming, as a working hypothesis that a hot deconfined medium is formed) is then simulated through the Langevin equation described at length in our previous works~\cite{Alberico:2011zy,Alberico:2013bza,Beraudo:2014boa}.
The latter requires the knowledge of the transport coefficients of the heavy quarks in the medium. 
As in our previous studies, we choose the ones provided by weak-coupling~\cite{Alberico:2011zy,Alberico:2013bza} and lattice-QCD calculations~\cite{Banerjee:2011ra,Francis:2011gc,Francis:2013cva,Francis:2015daa} and compare the predictions obtained in the two different scenarios.
Weak coupling calculations are performed by separating hard (treated in pQCD) and soft (including HTL resummation of medium effects) collisions. Lattice-QCD results, on the other hand, refer to a non perturbative setup, in which however the quark is treated as a static colour source; hence the kinematic range in which we can rely on a solid first-principle calculation in a non-perturbative domain is quite limited, being restricted to small quark velocities. 
The heavy-quark stochastic dynamics is followed until they reach a fluid-cell below a decoupling temperature $T_d$ (in this study set to 155 MeV), where they are made hadronize. The kinematics of the charm and beauty hadrons is defined at this stage and their possible rescatterings in the hadronic phase are neglected.

To describe hadronization in the presence of a hot medium we adopt the model developed in a previous study by us~\cite{Beraudo:2014boa}, to which we refer the reader for its detailed description; here, we just summarize its main features.
In the fluid-cell reached by the heavy-quark $Q$ one extracts a light antiquark $\overline{q}_{\rm light}$ (up, down or strange, with relative thermal abundances dictated by the ratio $m/T_{\rm dec}$) from a thermal momentum distribution corresponding to the temperature $T_{\rm dec}$ in the Local Rest Frame (LRF) of the fluid; information on the local fluid four-velocity $u^\mu_{\rm fluid}$ provided by hydrodynamics allows one to boost the momentum of $\overline{q}_{\rm light}$ from the LRF to the laboratory frame. 
A string is then constructed joining the endpoints given by $Q$ and $\overline{q}_{\rm light}$ and is then passed to PYTHIA 6.4~\cite{Sjostrand:2006za} to simulate its fragmentation into hadrons (and their final decays). In agreement with PYTHIA, in evaluating their momentum distribution, light quarks are taken as ``dressed'' particles with the effective masses $m_{u/d}\!=\!0.33$ GeV and $m_s\!=\!0.5$ GeV.

In A-A collisions the above hadronization mechanism turns out to provide a better agreement between the results of our model and the experimental data, with respect to the employment of standard in-vacuum fragmentation functions~\cite{Beraudo:2014boa}. The collective motion inherited from the light thermal parton increases the radial and elliptic flow of the final charmed and beauty hadrons, whose momentum and angular distributions display features closer to the actual experimental findings, such as the bump in the D-meson $R_{\rm AA}$ at low-$p_T$ ($p_T\!\approx\!1.5$ GeV/c) measured by STAR in Au-Au collisions at RHIC~\cite{Adamczyk:2014uip} and the sizable $v_2$ observed by ALICE in Pb-Pb collisions at the LHC~\cite{Abelev:2013lca}.
A few words are in order concerning the relationship between the model here adopted and the standard coalescence calculations. Both of them involve some mechanism of recombination with light medium particles, which transfer their collective motion to the heavy-flavour hadrons, giving rise overall to a more satisfactory agreement with the experimental findings compared to vacuum fragmentation. 
Coalescence is a $2\to 1$ process, occurring with high probability when the wave-function of the final D meson (if one considers charm) and the wave-packets of the two partons (the $c$ quark and a light antiquark from the medium) display a sizable overlap; this happens when the initial partons are sufficiently close in space and have comparable velocities. The effect of heavy-quark coalescence with light partons on the final particle spectra was studied in detail for instance in~\cite{vanHees:2007me}, where it turned out to provide a better description of the $R_{\rm AA}$ and $v_2$ of heavy-flavour decay electrons at RHIC.
Its relevance for possible modifications of the heavy-flavour hadrochemistry, such as an enhanced production of $D_s$ mesons, was also pointed out in~\cite{He:2012df}.
Our hadronization model, on the other hand, is based on a multistep $2\to 1\to N$ mechanism. One first combines a $Q$ with a thermal $\overline{q}$, independently of its kinematics, giving rise to a string of invariant mass $M$, which eventually decays into $N$ hadrons through excitation of $q\overline{q}$ pairs from the vacuum.

In the following section we check whether the above setup, which assumes the formation of a hot deconfined medium (albeit of small size) affecting the propagation and hadronization of heavy quarks even in the collision of small systems such as d-Au and p-Pb, is able to provide a consistent description of the current experimental data.

\section{Results: D/B-mesons and HF electrons}
\begin{figure}[!t]
\begin{center}
\includegraphics[clip,width=0.48\textwidth]{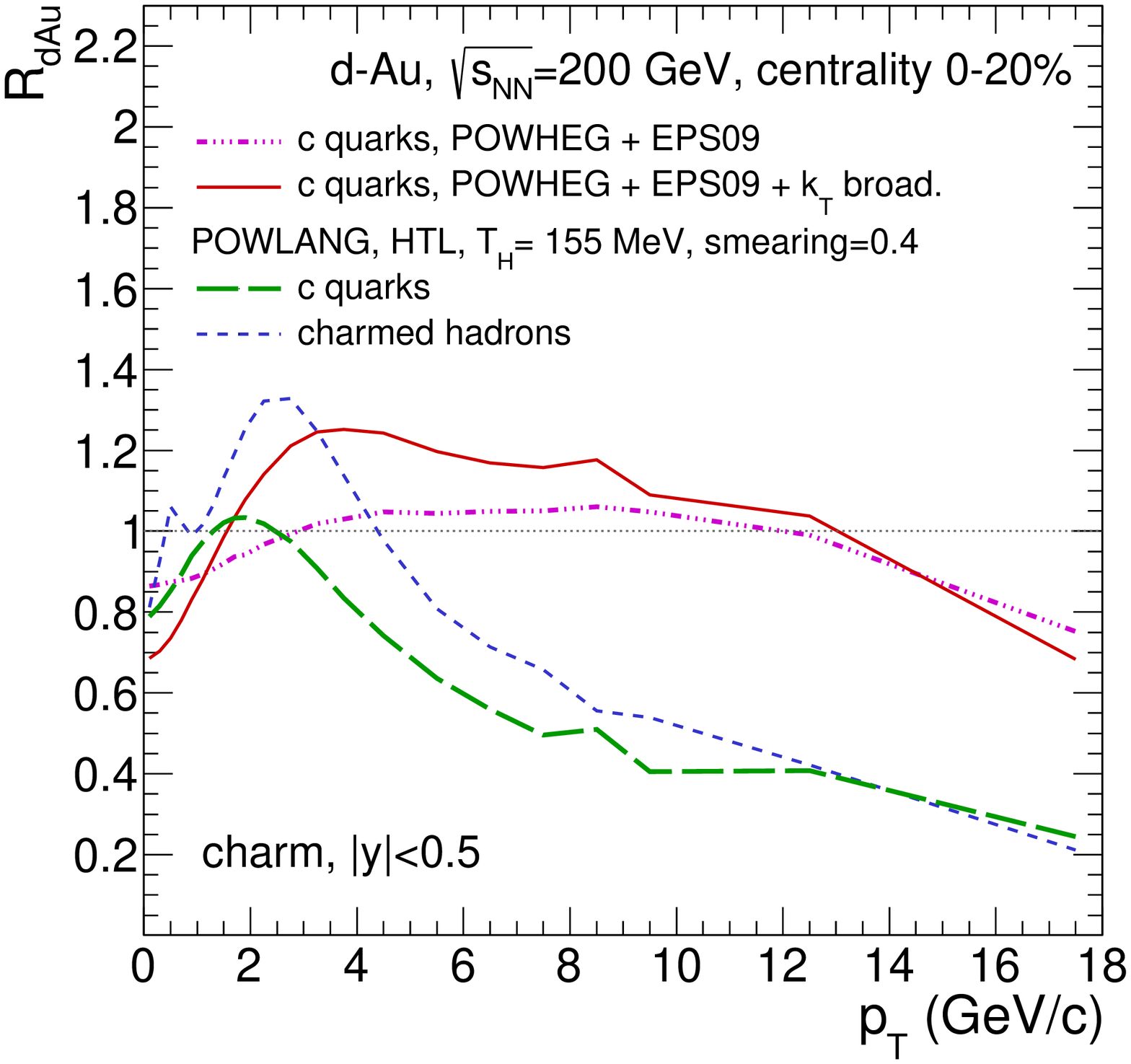}
\includegraphics[clip,width=0.48\textwidth]{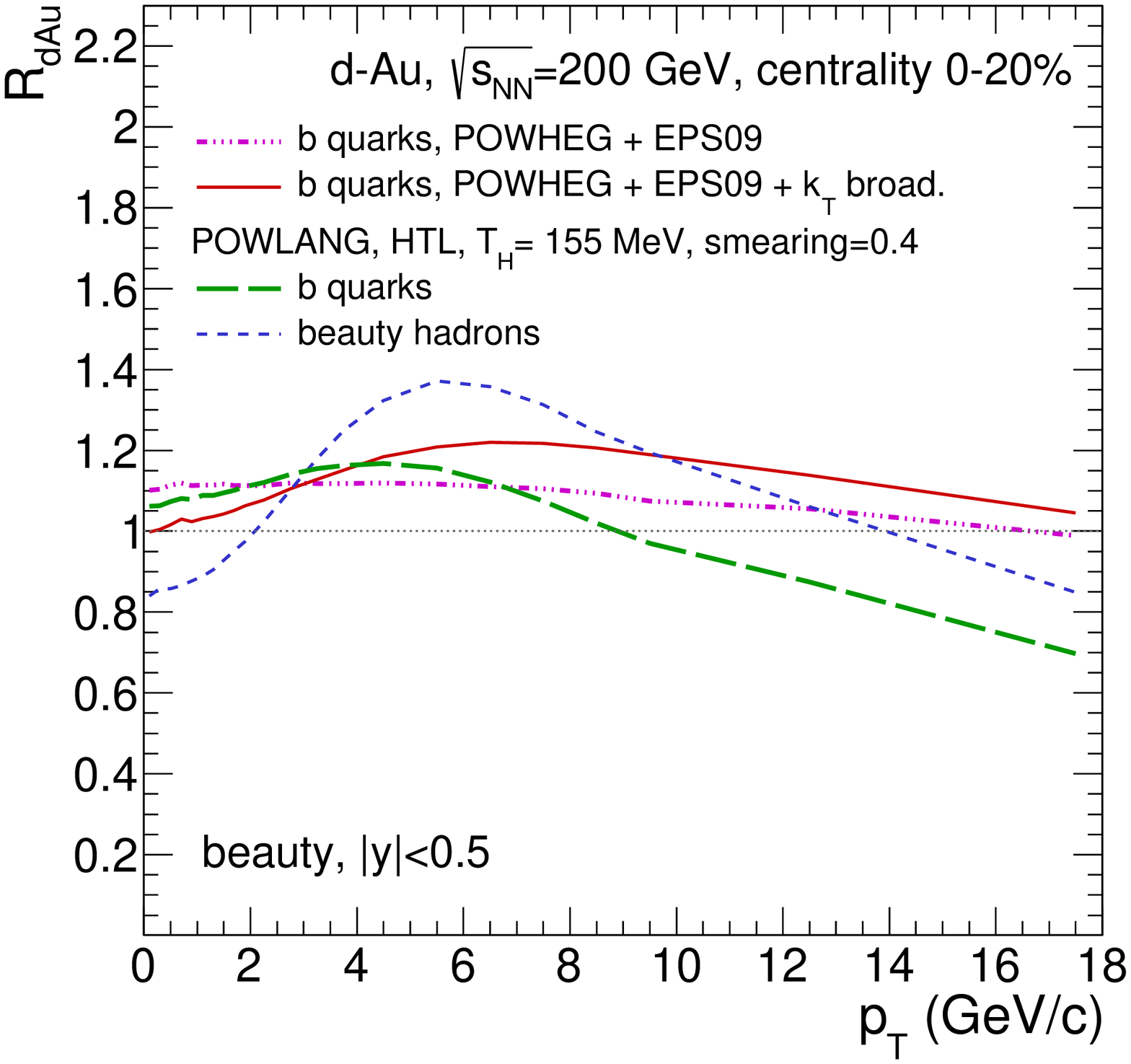}
\caption{The charm and beauty $R_{\rm dAu}$ in d-Au collisions at RHIC at the quark and hadron level. Results including only initial-state effects (nPDF's and $k_T$ broadening) are compared to the ones supplemented with the Langevin evolution in the plasma.}\label{fig:RdAU_qh} 
\end{center}
\end{figure}
\begin{figure}[!t]
\begin{center}
\includegraphics[clip,width=0.48\textwidth]{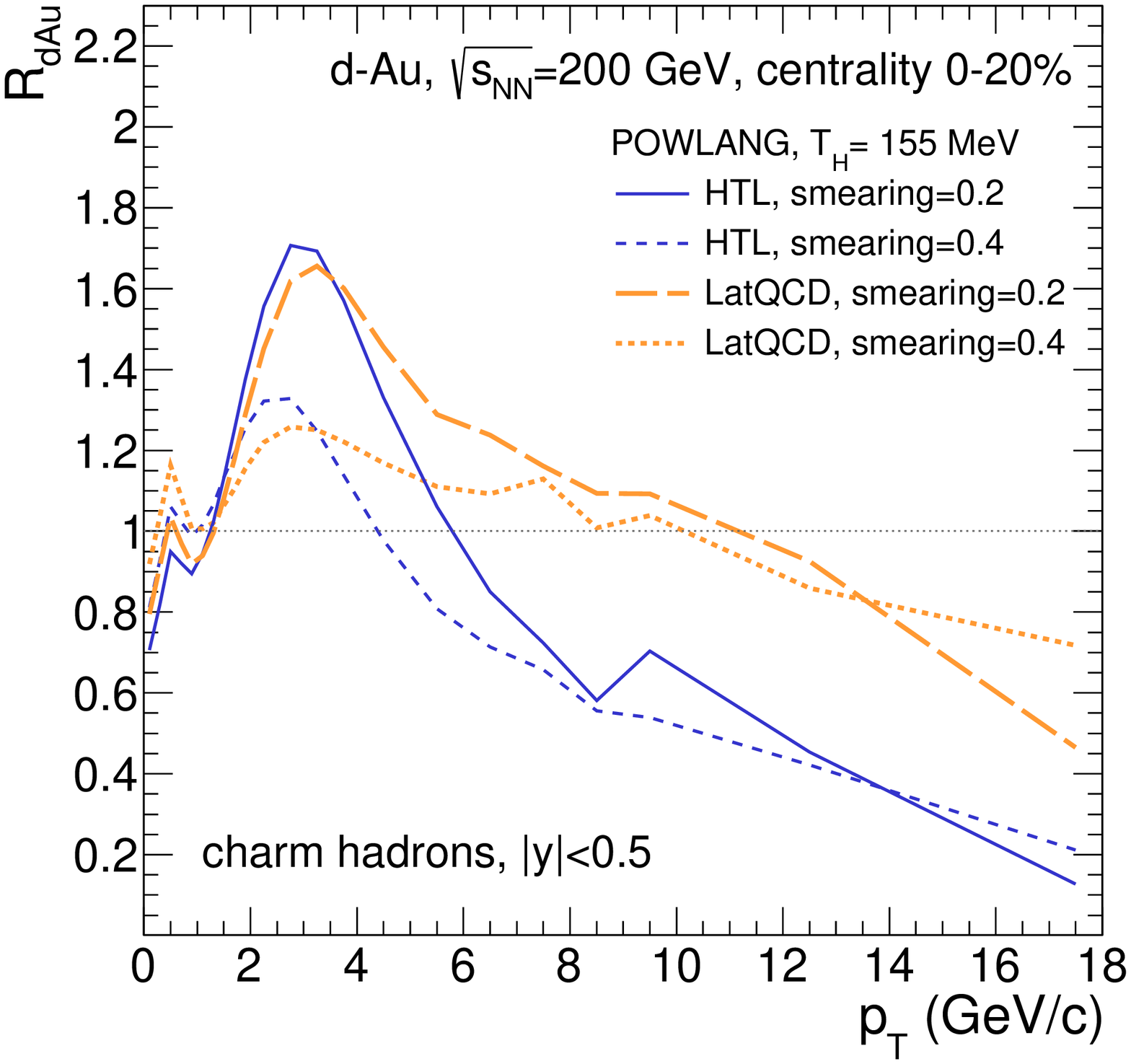}
\includegraphics[clip,width=0.48\textwidth]{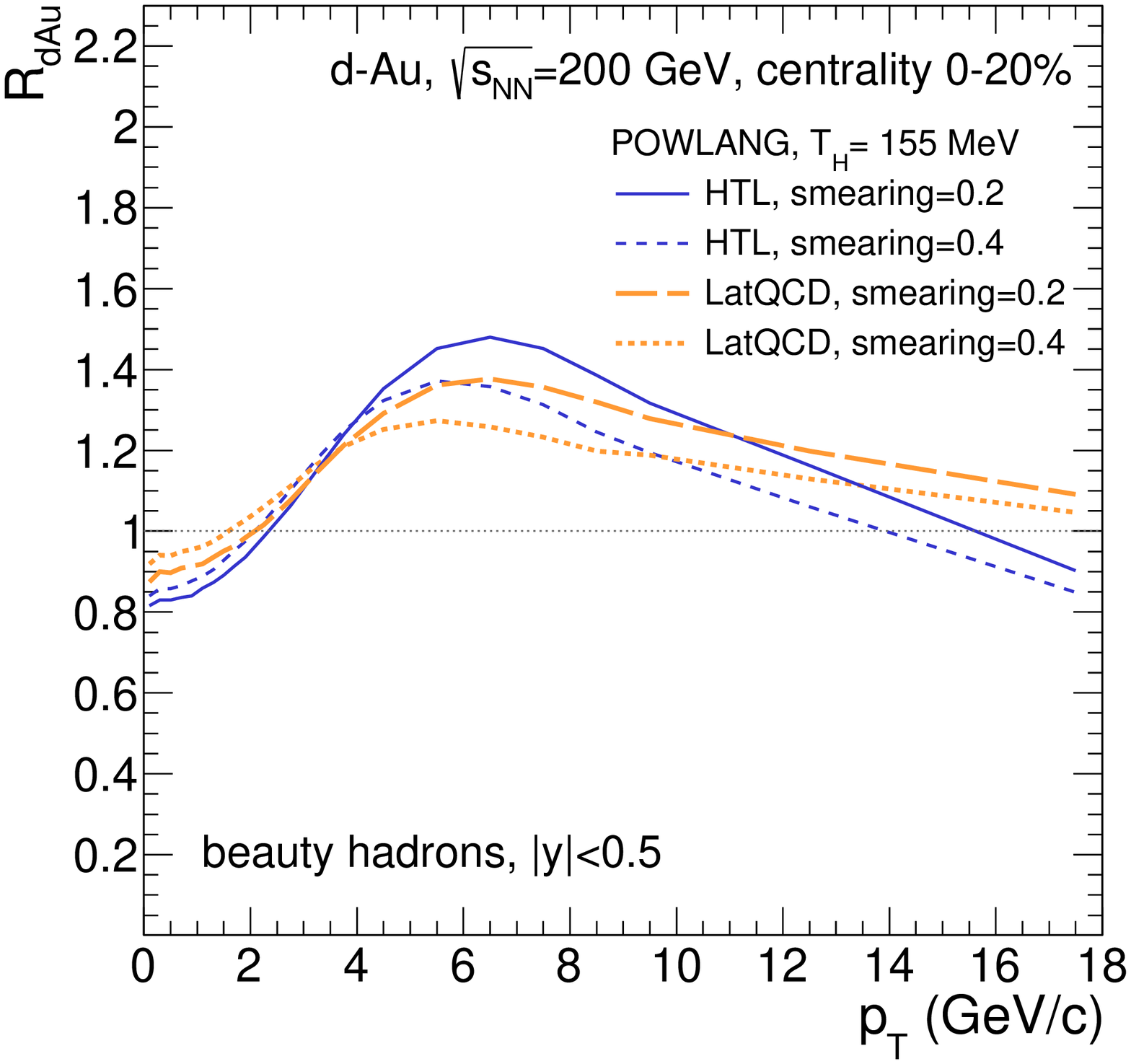}
\caption{The nuclear modification factor of charm and beauty hadrons in d-Au collisions at $\sqrt{s_{\rm NN}}\!=\!200$ GeV. Results of Langevin simulations with weak-coupling and lattice-QCD transport coefficients and different values of the initial smearing are compared.}\label{fig:RdAU_h_HTLvsLat} 
\end{center}
\end{figure}
In this section we display the results obtained with our transport setup (referred to as POWLANG, as for the A-A case) for the production of HF hadrons and decay-electrons in d-Au and p-Pb collisions at RHIC and LHC center-of-mass energies, respectively. We will compare them to the currently available experimental data obtained by the PHENIX and ALICE collaborations.
Actually, also CMS~\cite{Khachatryan:2015uja} has recently obtained results for the B-meson $R_{\rm pPb}$, but so far limited to too high-$p_T$ to make a comparison with an approach based on the Langevin equation meaningful.

The first experimental hints of possible final-state effects affecting heavy-flavour production in small systems were provided by the PHENIX results on the nuclear modification factor of non-photonic electrons in central d-Au collisions at $\sqrt{s_{\rm NN}}\!=\!200$ GeV~\cite{Adare:2012yxa}, with central values around $R_{\rm dAu}\!\approx\! 1.4$ over a quite extended $p_T$-range, from 1 to 5 GeV/c. Indeed, since -- in the light of A-A results -- one tended to associate a possible in-medium interaction with a quenching of the spectrum (i.e. $R_{\rm dAu}\!<\!1$) and since the systematic uncertainties from the background subtraction were large, people did not give at the beginning the proper importance to these results, considering them compatible with no medium-effect.
A study was actually carried out~\cite{Sickles:2013yna}, showing that, employing blast-wave spectra for D and B-mesons with parameters fixed by light hadron spectra (i.e. assuming that they share a common flow with an expanding medium), one would have been able to explain the enhanced production of NPE's in the moderate-$p_T$ domain analyzed. Such a picture, albeit interesting, lacks a microscopic dynamical justification.
Here we wish then to display the findings of the transport setup presented in the previous section, checking whether the combined effect of Langevin dynamics in the QGP and in-medium hadronization can lead to results in agreement with the experimental data. 

We start considering d-Au collisions at RHIC, focusing on the 20\% most central events, for which experimental data are available. 
In Fig.~\ref{fig:RdAU_qh} we show how the formation of a hot deconfined medium in the collision affects the HF quark and hadron spectra, by modifying their propagation and subsequent hadronization. The nuclear modification factor is characterized by a bump at intermediate $p_T$ (smaller for the quarks and more pronounced for the hadrons) that we attribute to radial flow; in particular, the larger effect at the hadron level is due to the additional flow inherited from the light quarks (the recombination and subsequent string fragmentation can of course slightly smear also the final hadron rapidity with respect to the one of the parent heavy quark). The curves display the results of calculations performed with weak-coupling heavy-quark transport coefficients, characterized by a steep increase with the particle momentum, hence the sizable quenching of charm at high $p_T$, shown in the figure up to a kinematic region out of the domain of validity of a Langevin picture. We also show the curves containing only cold nuclear matter(CNM) effects (nPDF's and $k_T$-broadening). Notice how the effect of the Cronin broadening is quite large at intermediate $p_T$, but, in our calculations, is washed out by the subsequent energy-loss in the deconfined medium. Therefore, the bump in the nuclear modification factor at the hadron level comes from the interplay of several effects, besides the CNM ones: low-$p_T$ charm quarks are pushed to higher momenta by their scatterings in the expanding medium, high-$p_T$ quarks tend to lose part of their energy and, finally, at hadronization the light partons transfers to the charmed hadrons part of their radial flow.  

In Fig.~\ref{fig:RdAU_h_HTLvsLat} we display the predictions of our transport setup at the hadron level, both for charm and beauty, exploring different choices of the smearing of the initial entropy-density in the transverse plane and of the transport coefficients, from weak-coupling and lattice-QCD calculations. In the last case no velocity-dependence of the heavy-quark momentum-diffusion coefficient (evaluated on the lattice for the case of a static quark) was assumed, which explains why, at large $p_T$, results stay close to unity. The important phenomenological outcome, however, is that, independently of the choice of the transport coefficients, all the curves display a bump around the same $p_T$ values, attributed -- within our setup -- to radial flow.

\begin{figure}[!h]
\begin{center}
\includegraphics[clip,width=0.48\textwidth]{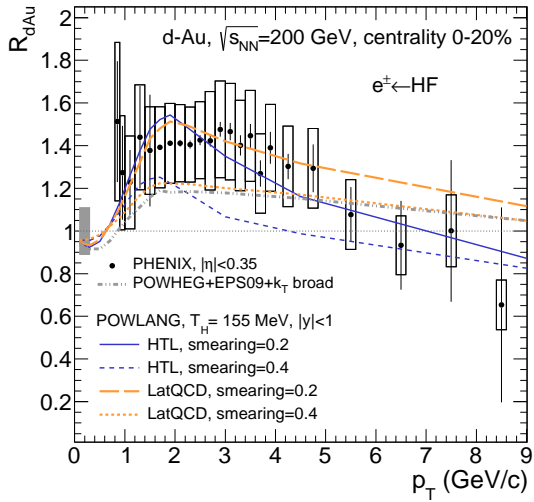}
\caption{The nuclear modification factor of HF decay electrons in d-Au collisions at $\sqrt{s_{\rm NN}}\!=\!200$ GeV. Results of Langevin simulation with weak-coupling and lattice-QCD transport coefficients and different values of the smearing parameter in the Glauber-MC initialization are shown and compared to PHENIX data~\cite{Adare:2012yxa}. {For comparison, we also display, in grey, the curve including only CNM effects (nPDF's and $k_T$-broadening)}.}\label{fig:RdAU_e} 
\end{center}
\end{figure}
Finally, we let HF hadrons decay semi-leptonically and we study the resulting electron spectra. Results for the nuclear modification factor $R_{\rm dAu}$ of HF decay electrons, corresponding to different transport coefficients and Glauber-MC initialization, are collected in Fig.~\ref{fig:RdAU_e}. Due to the large systematic uncertainties all curves look compatible with the data, which are not able to discriminate among the various choices of parameters. However, they are sufficient to see that the enhancement in the $R_{\rm dAu}$ of NPE's at moderate $p_T$ can be accommodated by models which, on top of cold nuclear matter effects, include also a stage of partonic transport in the hot plasma accompanied by in-medium hadronization. Therefore, within such a framework, the enhancement of the $R_{\rm dAu}$ of HF decay electrons reflects the radial flow acquired by the parent D and B mesons. {We also show in grey the result obtained including only CNM effects (nuclear modification of the PDF's and $k_T$-broadening) followed by in-vacuum independent fragmentation, which tends to slightly undershoot the data in the intermediate $p_T$ region. Notice that the fact that difference with the POWLANG curves is not dramatic is not due to the absence of final-state medium effects in the last case, but to the combined effect of parton energy-loss (tending to quench the spectra at high $p_T$) and of in-medium hadronization (responsible for most of the final HF radial-flow): this could be already inferred from the curves displayed in Fig.~\ref{fig:RdAU_qh}.}

\begin{figure}[!h]
\begin{center}
\includegraphics[clip,width=0.48\textwidth]{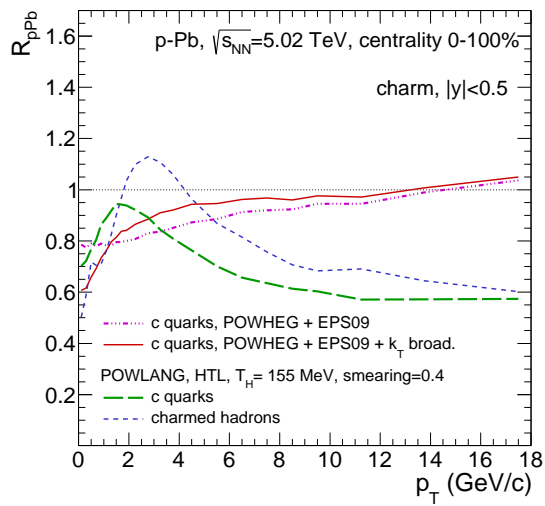}
\caption{The nuclear modification factor of charmed quarks and hadrons in 0-100\% p-Pb collisions at $\sqrt{s_{\rm NN}}\!=\!5.02$ TeV. One can appreciate how a significant fraction of radial flow is actually acquired at hadronization, from the recombination with light thermal partons. Outcomes of Langevin calculations refer to weak-coupling transport coefficients. We also show the curves including only CNM effects.}\label{fig:LHC_cvsD_RAA} 
\end{center}
\end{figure}
\begin{figure}[!t]
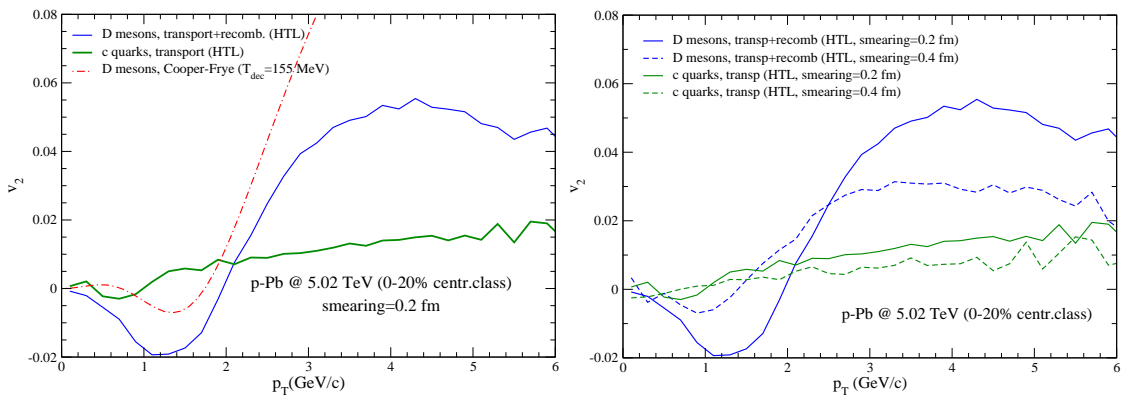

\begin{center}
\includegraphics[clip,width=0.48\textwidth]{v2_cDtherm_col.eps}
\includegraphics[clip,width=0.48\textwidth]{v2_Dvsc_syst.eps}
\caption{The elliptic-flow of charmed quarks and hadrons in the 0-20\% most central p-Pb collisions at $\sqrt{s_{\rm NN}}\!=\!5.02$ TeV. The curves refer to weak-coupling transport coefficients and various values of the initial smearing. The major contribution to the final $v_2$ comes from hadronization, due to the recombination with light thermal partons flowing with the medium. The D-meson $v_2$ at low-$p_T$ looks qualitatively similar to the one from a Cooper-Frye decoupling.}\label{fig:LHC_cvsD_v2} 
\end{center}
\end{figure}
We now move to p-Pb collisions at the LHC, at $\sqrt{s_{\rm NN}}\!=\!5.02$ TeV, where first experimental data for D-mesons, high-$p_T$ B-mesons, and HF decay electrons are available. Before addressing a systematic comparison with the accessible observables we wish to study separately, also in this case, the effects of the transport in the QGP phase and of in-medium hadronization.
In Figs.~\ref{fig:LHC_cvsD_RAA} and~\ref{fig:LHC_cvsD_v2} we show our model predictions for the nuclear modification factor (integrated over centrality, i.e. 0-100\%) and elliptic flow (in the 0-20\% centrality class) of charmed quarks and hadrons. Results refer to weak-coupling transport coefficients. As one can see, part of the final radial and elliptic flow of the charmed hadrons is actually acquired at hadronization, via the recombination with light thermal partons which participate in the collective motion of the medium. The effect is particularly evident in the case of $v_2$, which is small at the level of charm quarks at decoupling and reaches values around 5\% for D mesons; interestingly, in this case the $p_T$ dependence looks similar to the one from a Cooper-Frye decoupling of thermalized particles. 

\begin{figure}[!h]
\begin{center}
\includegraphics[clip,width=0.48\textwidth]{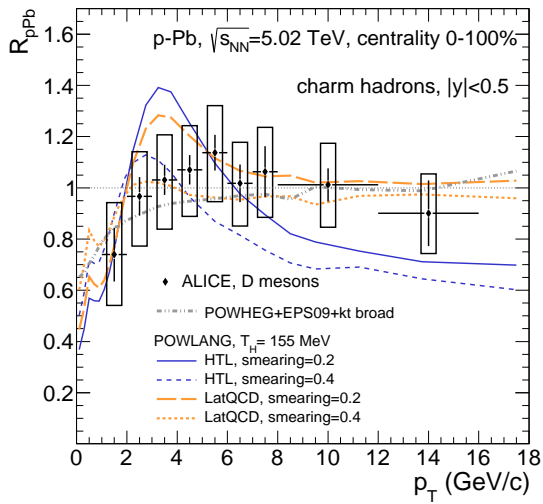}
\caption{The nuclear modification factor of charmed hadrons in 0-100\% p-Pb collisions at $\sqrt{s_{\rm NN}}\!=\!5.02$ TeV at the LHC. POWLANG results with HTL and l-QCD transport coefficients and different Glauber-MC initial conditions are compared to ALICE data for the D-meson $R_{\rm pPb}$~\cite{Abelev:2014hha}. {For comparison, we also display, in grey, the curve including only CNM effects (nPDF's and $k_T$-broadening)}.}\label{fig:LHC_RpPb_D} 
\end{center}
\end{figure}
We now want to compare the outcomes of our transport calculations with the currently available experimental data. We start from the case of charmed hadrons, whose nuclear modification factor is shown in Fig.~\ref{fig:LHC_RpPb_D} after averaging over all D-meson species, since our model is not conceived to address changes in the HF hadrochemistry, which might affect exclusive measurements. POWLANG results with HTL and l-QCD transport coefficients and different values of the smearing in the Glauber-MC initial conditions are compared to ALICE data for the D-meson $R_{\rm pPb}$ in the 0-100\% centrality class, resulting from the average of $D^0$, $D^+$ and $D^{*+}$. Predictions obtained with different transport coefficients differ sizably at high-$p_T$, due to their different dependence on the quark momentum; notice that, in particular for charm, this is a region out of the domain of validity of a Langevin approach. However, all the explored cases display some qualitative similarities: model results are characterized by a bump around $p_T\!\approx3\!$ GeV/c (interpreted, as usual, as a signature of radial flow), accompained by a depletion at low-$p_T$, partially arising from the nPDF's (shadowing) and partially from the conservation of the total number of charmed particles, moved by the rescatterings to larger $p_T$. {Also shown, in grey, is the case including only CNM effects (nuclear PDF's and $k_T$-broadening), characterized by a monotonic rise of the nuclear modification factor with increasing $p_T$.}

\begin{figure}[!t]
\begin{center}
\includegraphics[clip,width=0.48\textwidth]{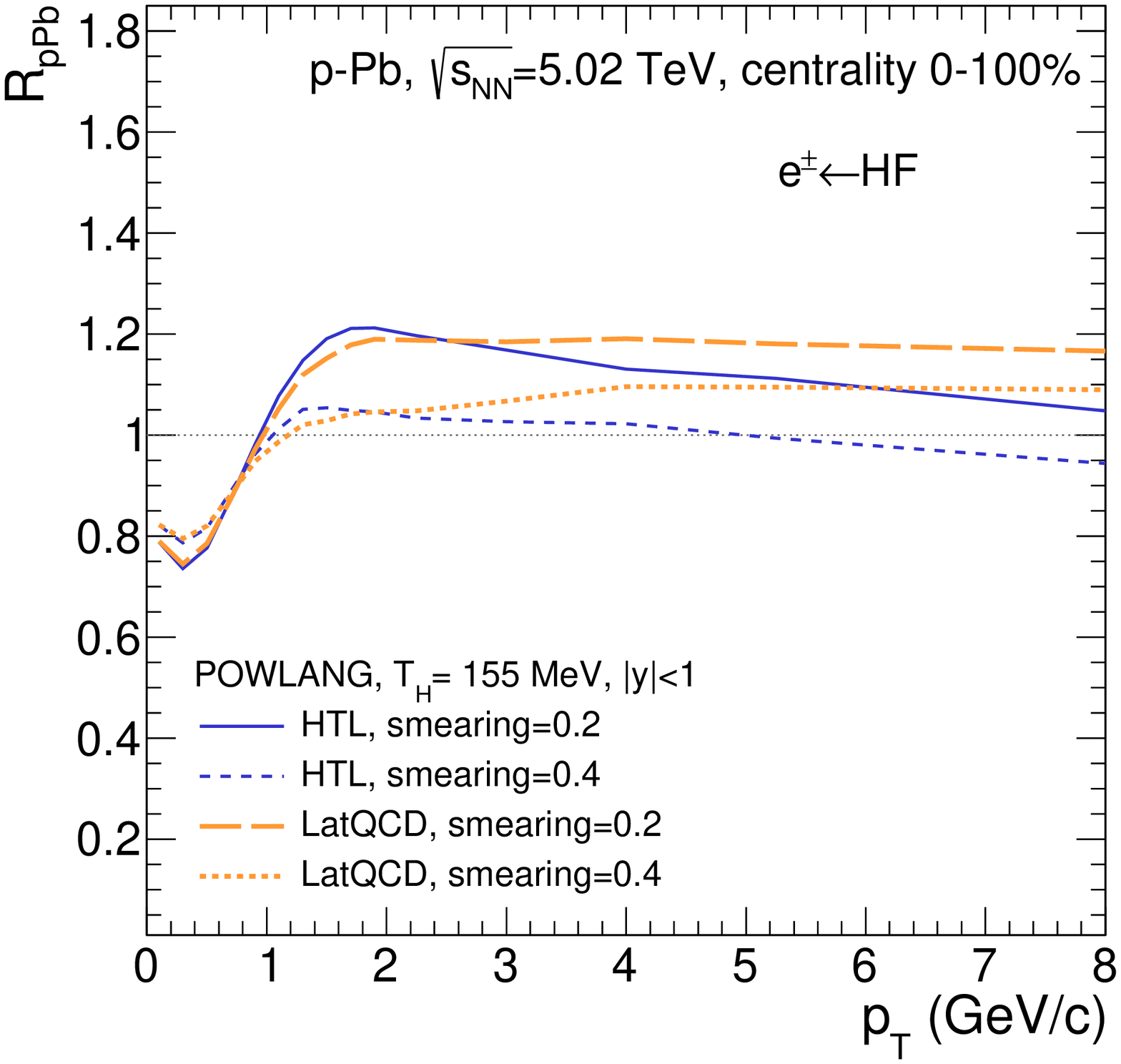}
\includegraphics[clip,width=0.48\textwidth]{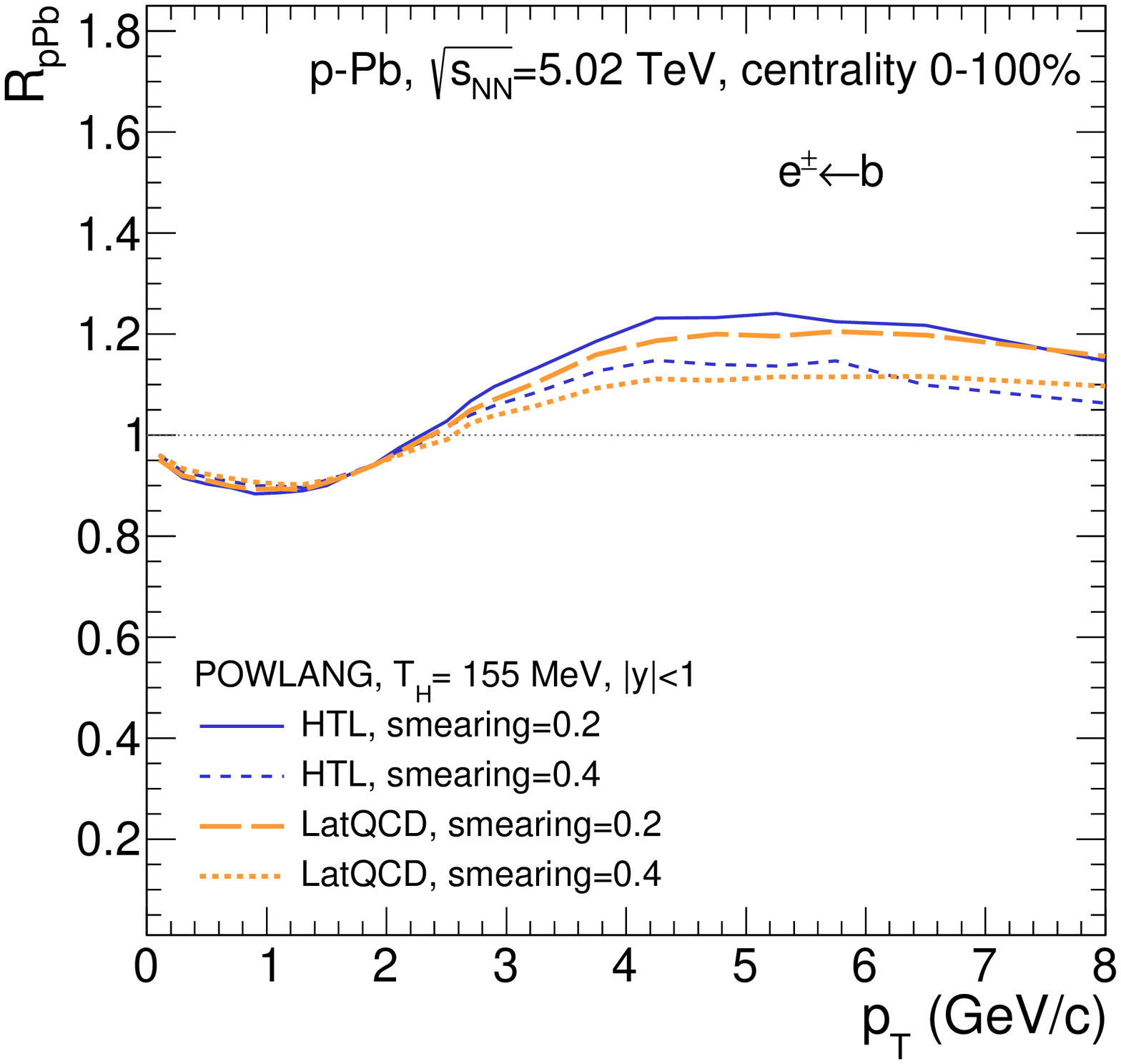}
\caption{The nuclear modification factor of HF electrons from charm and beauty decays (left panel) in p-Pb collisions at $\sqrt{s_{\rm NN}}\!=\!5.02$ TeV at the LHC. POWLANG results with HTL and l-QCD transport coefficients and different Glauber-MC initial conditions are shown. In the right panel model predictions for the beauty contribution are displayed.}\label{fig:LHC_RpPb_ecb} 
\end{center}
\end{figure}
Finally, in Fig.~\ref{fig:LHC_RpPb_ecb} we display the outcomes of our transport calculations for the spectra of electrons from charm and beauty decays $e_c\!+\!e_b$ (left panel), together with the corresponding results for the isolated $e_b$ contribution (right panel). As for the previous cases, theory calculations are characterized by a systematic uncertainty from the different possible choices of initialization and transport coefficients. However, overall, they provide indications for an $R_{\rm pPb}\gsim 1$ over a quite extended $p_T$-range, in qualitative agreement with the current experimental data~\cite{Adam:2015qda}, also affected by large systematic error bars.

\section{Conclusions and perspectives}
In this paper we have tried to answer the question whether the possible formation of a hot deconfined medium in the collisions of small systems, like d-Au or p-Pb, suggested by several soft observables involving light hadrons, may affect also the momentum and angular distributions of heavy-flavour particles. We have shown that transport calculations, accounting for medium modifications of both the propagation and the hadronization of heavy quarks, lead to results in qualitative and quantitative agreement with the current experimental data, some of which, on the contrary, display some tension when compared to approaches including simply initial-state effects, like nPDF's. 
The results we got lead one to look at HF observables in small systems from a conceptually new point of view with respect to the traditional idea of considering p(d)-A collisions just as a tool to estimate cold nuclear-matter effects.

Medium effects on heavy-flavour production in small systems have been recently studied also by other authors. The approach followed in~\cite{Xu:2015iha} is close to the one adopted for our analysis, namely a Langevin-like dynamics for the heavy quarks on top of a proper hydrodynamic evolution of the medium.  
On the other hand the perspective employed in~\cite{Kang:2014hha} is quite different: modifications of the HF $p_T$-spectra in p(d)-A collisions are attributed, in the A-going direction, to higher-twist effects such as incoherent multiple (double) parton scattering in the nuclear medium. Furthermore, ALICE data for the D-meson $R_{\rm pPb}$ at mid-rapidity turned out to be well described, within the current large systematic error-bars, also by the model of Ref.~\cite{Sharma:2009hn}, based on momentum broadening and energy-loss in cold nuclear matter. 

The growing experimental and theoretical interest on the subject and the first results we obtained encourage us to perform more detailed studies, requiring a much heavier numerical effort, like full hydro+transport event-by-event simulations, with a more realistic (3+1)D hydrodynamic background, in order to access a wider set of observables in a larger rapidity-range. This will also allow one to discriminate among our predictions and the ones of the other models, which don't assume the formation of a hot medium, but involve only cold nuclear matter effects~\cite{Kang:2014hha,Sharma:2009hn,Fujii:2013yja,Fujii:2015lld}. We leave such an issue for future publications.  

\appendix

\section{Initial conditions: setting the parameters}\label{app:init}
The parameters necessary to specify the Glauber-MC initial conditions for the hydrodynamic evolution in the d-Au and p-Pb cases, in particular the constant $K$ (having dimensions of an inverse length) providing the entropy released by each nucleon-nucleon inelastic interaction, are fixed by matching the results of the optical and Monte-Carlo versions of the Glauber model in A-A collisions.

We proceed as follows.
We assume the initial entropy density in a nucleus-nucleus collision at impact parameter $b$ to be given, within the optical Glauber model, by
\beq
s(\x,b)=s_0\frac{n_{\rm coll}(\x,b)}{n_{\rm coll}(\x\!=\!0,b\!=\!0)},
\eeq    
where $s_0$ (in fm$^{-3}$) sets its maximum value at the origin. Other choices are of course legitimate; we took this $n_{\rm coll}$ scaling for consistency with our previous A-A studies, based on the Glauber-model initialization for the background medium of Ref.~\cite{Luzum:2008cw,Luzum:2009sb}.
The entropy per unit (space) rapidity ($\tau_0$ being the thermalization time) in the centrality-class $\cal{C}_1-\cal{C}_2$ is obtained integrating over the transverse plane and averaging over the impact parameter:
\beq
\frac{1}{\tau_0}\left\langle\frac{dS}{d\eta_s}\right\rangle_{\cal{C}_1-\cal{C}_2}^{\rm opt}\!\!\!=s_0\frac{\langle N_{\rm coll}\rangle_{\cal{C}_1-\cal{C}_2}^{\rm opt}}{n_{\rm coll}(\x\!=\!0,b\!=\!0)}.\label{eq:dSdetaopt}
\eeq
In the optical-Glauber approach $\langle N_{\rm coll}\rangle_{\cal{C}_1-\cal{C}_2}^{\rm opt}$ is given by:
\beq
{\langle N_{\rm coll}\rangle_{ {\cal C}_1-{\cal C}_2}^{\rm opt}}\equiv\frac{\int_{b_1}^{b_2} bdb\,{N_{\rm coll}^{\rm in.evt}(b)}\, p_{\rm in}^{AB}(b)}{\int_{b_1}^{b_2} bdb\, p_{\rm in}^{AB}(b)}={\frac{\int_{b_1}^{b_2} bdb\, N_{\rm coll}(b)}{\int_{b_1}^{b_2} bdb\, p_{\rm in}^{AB}(b)}}.\label{eq:averNcoll}
\eeq
For the sake of completeness we remind the meaning of the various terms in the above (nuclear thickness and overlap functions are assumed normalized to 1):
\begin{itemize}
\item \emph{Probability of having {at least one} inelastic interaction}
\beq
{p_{\rm in}^{AB}(b)}=\sum_{{n=1}}^{AB}P(n,b)={1-[1-\widehat{T}_{AB}(b)\sigma_{\rm in}^{NN}]^{AB}}\nonumber
\eeq
\item Number of {binary collisions} ({per $A-B$ crossing}):
\beq
{N_{\rm coll}(b)}=\sum_{n=1}^{AB}n\,P(n,b)={AB\,\widehat{T}_{AB}(b)\,\sigma_{\rm in}^{NN}}\nonumber
\eeq
\item Number of {binary collisions} \emph{per inelastic event}:
\beq
{N_{\rm coll}^{\rm in.evt}(b)=N_{\rm coll}(b)}/{p_{\rm in}^{AB}(b)}\nonumber
\eeq
\end{itemize}
Centrality classes are defined as percentiles of the total geometric cross-section:
\beq
\frac{\int_0^{{b_{1/2}}} bdb\, p_{\rm in}^{AB}(b) }{\int_0^\infty bdb\, p_{\rm in}^{AB}(b) }=\frac{\sigma(b<b_{1/2})}{\sigma^{\rm tot}}
\eeq
In the Glauber-MC approach on the other hand the entropy-density in a given event is distributed in the transverse plane according to Eq.~(\ref{eq:sdens}). Its integral over the transverse plane is directly proportional to the number of binary collisions of the considered event. The average entropy per unit rapidity in a given centrality class is now given by:
\beq
\frac{1}{\tau_0}\left\langle\frac{dS}{d\eta_s}\right\rangle_{\cal{C}_1-\cal{C}_2}^{\rm MC}\!\!\!=K\langle N_{\rm coll}\rangle_{ {\cal C}_1-{\cal C}_2 }^{\rm MC},\label{eq:dSdetaMC}
\eeq
where now centrality classes are simply defined ordering the events according to $N_{\rm coll}$ and taking the various percentiles of the $(dN^{\rm evt}/dN_{\rm coll})$ distribution. By comparing Eqs.~(\ref{eq:dSdetaopt}) and (\ref{eq:dSdetaMC}) one can fix the parameter K (in fm$^{-1}$):
\beq
K=s_0\frac{\langle N_{\rm coll}\rangle_{\cal{C}_1-\cal{C}_2}^{\rm opt}/\langle N_{\rm coll}\rangle_{\cal{C}_1-\cal{C}_2}^{\rm MC} }{n_{\rm coll}(\x\!=\!0,b\!=\!0)}.\label{eq:matching}
\eeq

At RHIC we perform the matching considering the 0-5\% most central Au-Au collisions at $\sqrt{s_{\rm NN}}\!=\!200$ GeV; the result found for $K$ can be then directly used also in Glauber-MC simulations of d-Au collisions at the same center-of-mass energy. Setting, as done in our previous studies~\cite{Alberico:2011zy,Alberico:2013bza,Beraudo:2014boa}, $s_0\tau_0=84$ fm$^{-2}$ and evaluating the density and average number of binary collisions in the Glauber model one gets the value of the product $K\tau_0\!=\!3.99$ quoted in Table~\ref{tab:init}.

At the LHC the situation is different, since so far we don't have Pb-Pb and p-Pb collisions at the same center-of-mass energy. We have then to extrapolate the parameters fixed at $\sqrt{s_{\rm NN}}\!=\!2.76$ TeV to the 5.02 TeV case. We proceed then as follows. The initial entropy deposited by the colliding nuclei is assumed to be proportional to the rapidity density of the final charged particles produced in the collision. Considering for instance the percentile of the most central events one has 
\beq
\left.\frac{dN_{\rm ch}}{d\eta}\right|_{\rm centr}\!\!\!\sim\left\langle\frac{dS}{d\eta_S}\right\rangle_{\rm centr}\!\!\!=s_0\tau_0\frac{\langle N_{\rm coll}\rangle_{\rm centr }^{\rm opt}}{n_{\rm coll}(\x=0,b=0)}.\label{eq:entroestim}
\eeq
For the product $s_0\tau_0$ in Pb-Pb collisions at $\sqrt{s_{\rm NN}}\!=\!2.76$ TeV in our past studies~\cite{Alberico:2013bza,Beraudo:2014boa} we employed the value $s_0\tau_0=166.8$ fm$^{-2}$. In order to fix the parameter $K$ to employ in the Glauber-MC simulations at $\sqrt{s_{\rm NN}}\!=\!5.02$ TeV through Eq.~(\ref{eq:matching}) we need to estimate $dN_{\rm ch}/d\eta$, and then $s_0$, at such a center-of-mass energy for which so far we don't have experimental data in nucleus-nucleus collisions. Hence we must rely on extrapolation of experimental data at lower values of $\sqrt{s_{\rm NN}}$. In the case of A-A collisions the following empirical formula was found~\cite{Aamodt:2010pb}
\beq
\frac{1}{\langle N_{\rm part}\rangle/2}\frac{dN_{\rm ch}}{d\eta}\sim (\sqrt{s_{\rm NN}})^{0.3},\label{eq:sqrtdep}
\eeq
which nicely describe the evolution from RHIC to LHC, from $dN_{\rm ch}/d\eta\approx 660$ in the 0-6\% most central Au-Au events at $\sqrt{s_{\rm NN}}\!=\!200$ GeV~\cite{Back:2001ae,Back:2002uc} to $dN_{\rm ch}/d\eta\approx 1600$ in the 0-5\% centrality class for Pb-Pb collisions at $\sqrt{s_{\rm NN}}\!=\!2.76$ TeV~\cite{Aamodt:2010cz}. In extrapolating the initial entropy-density from $\sqrt{s_1}$ (2.76 TeV) to $\sqrt{s_2}$ (5.02 TeV) we will then employ the following formula, obtained combining Eqs.~(\ref{eq:entroestim}) and (\ref{eq:sqrtdep}):
\beq
(s_0\tau_0)_2=(s_0\tau_0)_2\frac{\langle N_{\rm part}\rangle_2}{\langle N_{\rm part}\rangle_1}\left(\frac{\sqrt{s_2}}{\sqrt{s_1}}\right)^{0.3}\displaystyle{\frac{\left(\frac{\langle N_{\rm coll}\rangle}{n_{\rm coll}(\0,0)}\right)_1}{\left(\frac{\langle N_{\rm coll}\rangle}{n_{\rm coll}(\0,0)}\right)_2}}.
\eeq
At the end we get
\beq
(s_0\tau_0)_{\sqrt{s}=5.02}\approx(s_0\tau_0)_{\sqrt{s}=2.76}\approx 200\,{\rm fm}^{-2},
\eeq
which leads to the estimate $K\tau_0\!=\!6.37$ quoted in Table~\ref{tab:init}.

\bibliography{paper_rev2}

\end{document}